\documentclass[%
superscriptaddress,
reprint,
 amsmath,amssymb,
prb,
]{revtex4-1}

\usepackage[utf8]{inputenc}
\usepackage[T1]{fontenc}

\usepackage{color}
\usepackage[pdftex]{graphicx}
\usepackage{graphicx}
\usepackage{dcolumn}
\usepackage{bm}
\usepackage{siunitx}
\usepackage{braket}
\usepackage{ulem}

\newcommand{\uu}{\ket{\uparrow\uparrow}} 
\newcommand{\ud}{\ket{\uparrow\downarrow}} 
\newcommand{\du}{\ket{\downarrow\uparrow}} 
\newcommand{\dd}{\ket{\downarrow\downarrow}} 

\newcommand{\Uuu}{\textrm{U}_{\alpha}} 
\newcommand{\Uud}{\textrm{U}_{\beta}} 

\newcommand{\txtind}[1]{_\text{#1}}
\newcommand{\txtexp}[1]{^\text{#1}}
\newcommand{\PS}{P_{\ket{\text{S}}}}

\begin{document}

\title{Distant spin entanglement via fast and coherent electron shuttling}

\author{Baptiste Jadot}
\email{baptiste.jadot@neel.cnrs.fr}
\affiliation{Univ. Grenoble Alpes, CNRS, Grenoble INP, Institut N\'eel, 38000 Grenoble, France}

\author{Pierre-Andr\'e Mortemousque}
\affiliation{Univ. Grenoble Alpes, CNRS, Grenoble INP, Institut N\'eel, 38000 Grenoble, France}

\author{Emmanuel Chanrion}
\affiliation{Univ. Grenoble Alpes, CNRS, Grenoble INP, Institut N\'eel, 38000 Grenoble, France}

\author{Vivien Thiney}
\affiliation{Univ. Grenoble Alpes, CNRS, Grenoble INP, Institut N\'eel, 38000 Grenoble, France}

\author{Arne Ludwig}
\affiliation{Lehrstuhl f{\"u}r Angewandte Festk{\"o}rperphysik, Ruhr-Universit{\"a}t Bochum, Universit{\"a}tsstra{\ss}e 150, D-44780 Bochum, Germany}

\author{Andreas D. Wieck}
\affiliation{Lehrstuhl f{\"u}r Angewandte Festk{\"o}rperphysik, Ruhr-Universit{\"a}t Bochum, Universit{\"a}tsstra{\ss}e 150, D-44780 Bochum, Germany}

\author{Matias Urdampilleta}
\affiliation{Univ. Grenoble Alpes, CNRS, Grenoble INP, Institut N\'eel, 38000 Grenoble, France}

\author{Christopher B{\"a}uerle}
\affiliation{Univ. Grenoble Alpes, CNRS, Grenoble INP, Institut N\'eel, 38000 Grenoble, France}

\author{Tristan Meunier}
\email{tristan.meunier@neel.cnrs.fr}
\affiliation{Univ. Grenoble Alpes, CNRS, Grenoble INP, Institut N\'eel, 38000 Grenoble, France}

\date{\today}

\begin{abstract}
In the quest for large-scale quantum computing, networked quantum computers offer a natural path towards scalability. Now that nearest neighbor entanglement has been demonstrated for electron spin qubits in semiconductors, on-chip long distance entanglement brings versatility to connect quantum core units. Here we realize the controlled and coherent transfer of a pair of entangled electron spins, and demonstrate their remote entanglement when separated by a distance of $\SI{6}{\micro m}$. Driven by coherent spin rotations induced by the electron displacement, high-contrast spin quantum interferences are observed and are a signature of the preservation of the entanglement all along the displacement procedure. This work opens the route towards fast on-chip deterministic interconnection of remote quantum bits in semiconductor quantum circuits.
\end{abstract}
\maketitle

\subsection*{Introduction}
Creating and manipulating entanglement among an assembly of qubits is a key ingredient for exploiting quantum parallelism in quantum computers. Demonstrating it at distance\cite{aspect_experimental_1981} has enabled quantum communication\cite{gisin_quantum_2002, pfaff_unconditional_2014, gao_observation_2012, imamoglu_quantum_1999} and quantum teleportation\cite{bouwmeester_experimental_1997, barrett_deterministic_2004, riebe_deterministic_2004} protocols where quantum states can be displaced at will in quantum circuits. This has been demonstrated in several quantum systems but remains a key functionality to be implemented for electron spins in semiconductor quantum circuits. Following the first nearest-neighbor entanglement demonstrations\cite{shulman_demonstration_2012,kandel_coherent_2019,veldhorst_two-qubit_2015,watson_programmable_2018}, research efforts focus on several distinct strategies to implement a long-range quantum mediator: coupling to a single microwave photon\cite{samkharadze_strong_2018, borjans_resonant_2020, landig_coherent_2018, viennot_coherent_2015}, long-range spin-spin interaction mediated through quantum dot systems\cite{baart_coherent_2017, malinowski_fast_2019} or controllable displacement of electron spins\cite{flentje_coherent_2017, fujita_coherent_2017, mortemousque_coherent_2018}. In this article, coherent electron shuttling is exploited as a source of remotely entangled electron spins.

In semiconductor circuits, two transfer approaches have been identified to preserve quantum information during the electron transfer. In a first strategy, the electron is displaced in a static array of quantum dots via a tunneling process. To protect the state coherence, the passage from one dot to another must be an adiabatic tunneling process. Coherent transfer over $\SI{5}{\micro m}$ has been demonstrated with a speed of approximately $\SI{100}{m/s}$, limited by the tunnel coupling between the dots\cite{flentje_coherent_2017, fujita_coherent_2017, mortemousque_coherent_2018}. The second strategy consists in shuttling electrons by moving the trapping potential along the channel. Efficient and fast electron transfer protocols between two distant quantum dots have been established using surface acoustic waves (SAW), with a speed of $\SI{2700}{m/s}$ in AlGaAs heterostructures\cite{mcneil_-demand_2011, hermelin_electrons_2011, takada_sound-driven_2019}.

In this article, we exploit the moving trapping potential associated with the propagation of a SAW to controllably separate and recombine two entangled electron spins on fast timescales. This is realized by the following three-step procedure: first, two electrons are prepared in a singlet spin state in the source dot. Second, the two electrons are sequentially transferred using a SAW. Finally, the electrons are recombined in the reception dot and single-shot spin readout is performed to evaluate the two-electron singlet state probability. To preserve and demonstrate long-distance entanglement, we rely on several important quantum functionalities, co-implemented in the same device: high-fidelity initialization and readout of two-electron spin states at both ends of the channel, a nanosecond-controlled electron separation and transfer procedure using SAW, and high-fidelity coherent rotations induced by the individual electron displacement. 

\begin{figure*}[!htp]
\includegraphics[width=180mm]{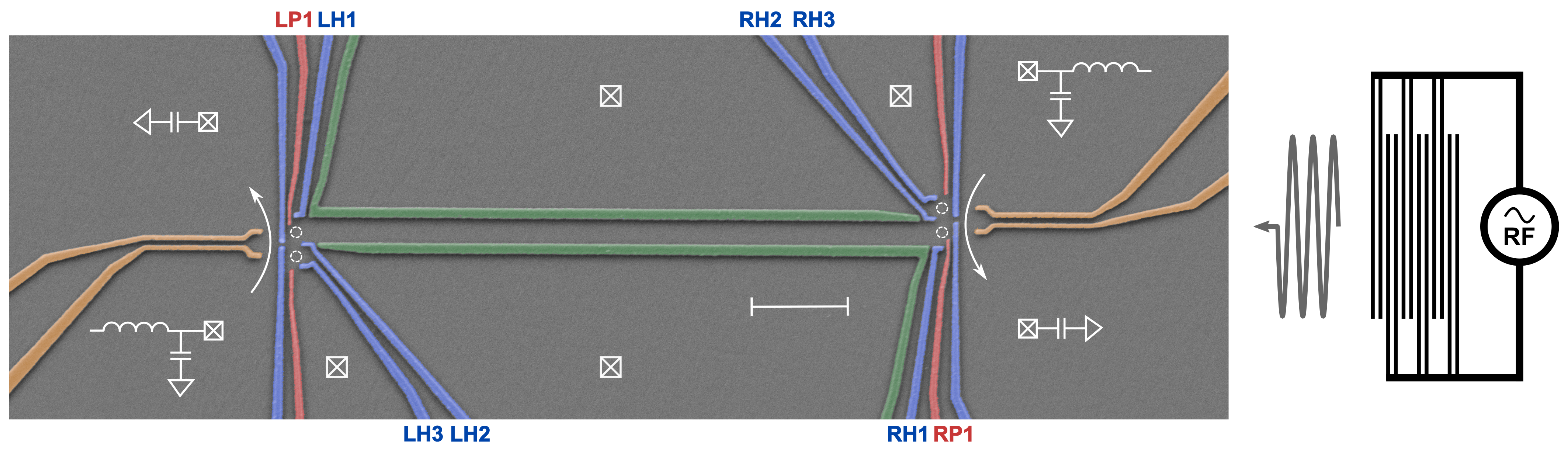}
\caption{ \textbf{Electron transfer protocol.}
False color scanning electron micrograph of an identical device. Two double quantum dots are connected by a $\SI{6}{\micro m}$ depleted channel (green gates), and probed by a local electrometer. The dotted circles represent the positions of the four quantum dots, and the scale bar is $\SI{1}{\micro m}$. Two electrons initially loaded in the source dot (right dot) are propelled towards the reception dot (left dot) with the help of a propagating sinusoidal electric field induced by a surface acoustic waves (SAW), located \SI{2}{mm} away on the right side of the structure.
}
\label{fig:sample}
\end{figure*}

\begin{figure*}[!htp]
\includegraphics[width=143mm]{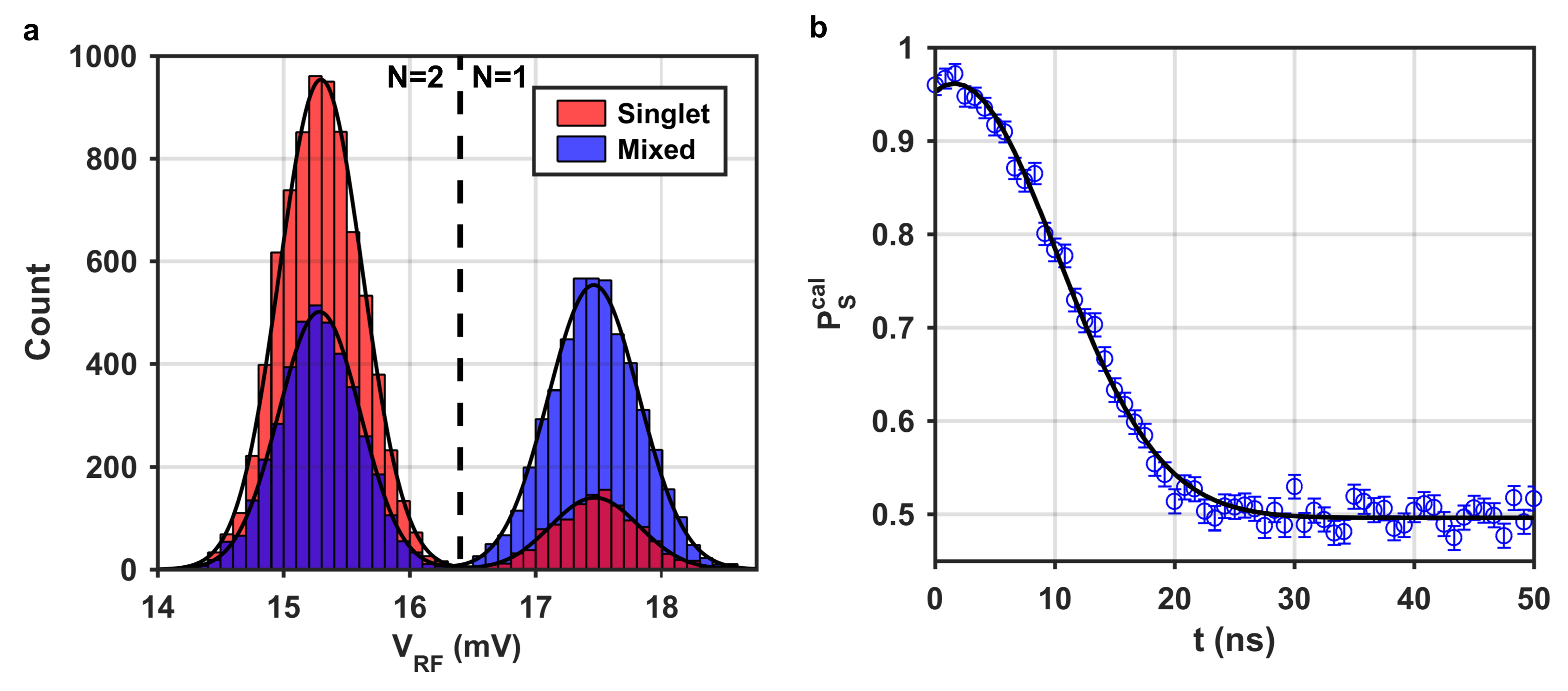}
\caption{ \textbf{Local spin manipulation.}
\textbf{a,} Histogram of the detector response after spin-to-charge conversion performed on the left double quantum dot, for two different spin initializations and for $\SI{200}{\micro s}$ integration time. To evaluate the two-electron spin state, a $\SI{16}{\micro s}$ voltage pulse increases abruptly the coupling to the reservoir, bringing the system in a configuration where a triplet spin state has a significantly greater probability to tunnel-out to the reservoir. After this pulse, the system is brought back to the isolated regime and charge is read-out for $\SI{200}{\micro s}$, leading to the assertion of a singlet spin state if two electrons remain (and a triplet state otherwise). The fidelity of this readout reaches \SI{95}{\%} in the reception dot, and has been calibrated according to Suppl. Inf. 3.
\textbf{b,} Time evolution of the two-spin mixing when the electrons are separated in  the reception double dot. The two electrons, initially prepared in a singlet spin state, are separated in adjacent dots for a few ns before recombination and spin readout. The resulting time evolution of the calibrated singlet probability is fitted with a Gaussian decay of characteristic time $T_2^* = \SI{12.2}{ns}$, imposed by the difference of nuclear environment between the two dots\cite{petta_coherent_2005}.
}
\label{fig:spin_manipulation}
\end{figure*}

\subsection*{Description of the experiment}
The sample, made from a GaAs/AlGaAs heterostructure, is represented in Fig. 1. The electrons, initially located on the right double quantum dot, are picked up by the propagating potential modulation and travel in moving quantum dots at the SAW velocity ($\SI{2700}{m/s}$)\cite{talyanskii_single-electron_1997, hermelin_electrons_2011, mcneil_-demand_2011}, completing the displacement across the channel in $t\txtind{S} = \SI{2.1}{ns}$. The injection and capture between static and moving dots are carried out in the so-called isolated regime in order to avoid any electron leakage to or from the reservoir (see Suppl. Inf. 2). On both sides of the nanostructure, efficient two-electron-spin initialization and measurement protocols are implemented. First, spin initialization in the singlet states reaches 0.95 fidelity when two electrons are loaded in the dot at $\si{\micro s}$ timescales. Second, the spin readout is based on the difference of the tunnel-rates towards the reservoir for the singlet and triplet spin states, as illustrated in Fig. 2a\cite{flentje_coherent_2017, hanson_spins_2007, meunier_nondestructive_2006}. The fidelity of the spin readout reaches a maximum of 0.95 in the reception dot (see Fig. 2a and Suppl. Inf. 3 for details). 

\begin{figure*}[!htp]
\includegraphics[width=138mm]{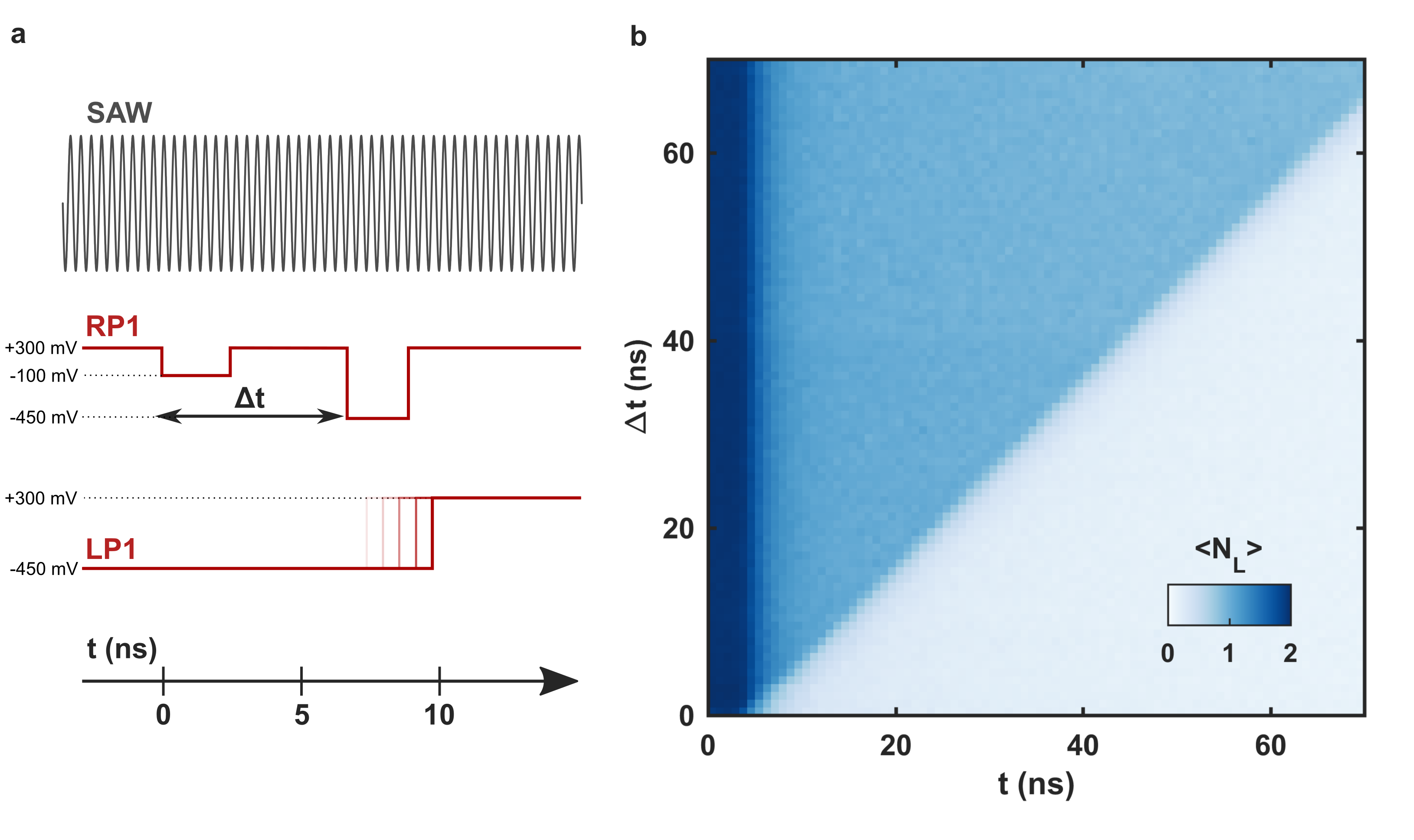}
\caption{ \textbf{Controllable injection in moving quantum dots.}
\textbf{a,} Scheme of the injection timing measurement. Two voltage pulses of a $\SI{2.5}{ns}$ duration, synchronized with the SAW burst, are applied on gate RP1 of the sending dot to trigger the injection of the two electrons into the moving quantum dots. To measure the electron distribution in the moving dots, the reception dot is kept in a configuration where electron catching is unlikely until a positive voltage pulse on LP1 is applied. We take the first electron injection as the reference time $t=\SI{0}{ns}$.
\textbf{b,} Average charge caught on the reception dot depending on the injection and catching pulse delays. Each pixel is the average of 4000 single-shot realizations. The vertical and diagonal boundaries correspond to the arrival time of the first and second electron, respectively. The vertical boundary begins at $t\txtind{S} = \SI{2.1}{ns}$ as a consequence of the time of flight at the SAW velocity of the first electron between the two static dots. Moreover, the clean and sharp diagonal boundary with unity slope confirm our ability to control the separation of two electrons by up to $\SI{70}{ns}$ with a nanosecond resolution, using a double AWG voltage pulse. The precision of the sending is measured close to a nanosecond and fixed by the rise time of the AWG.
}
\label{fig:delayed_sending}
\end{figure*}

\subsection*{Controlled injection into moving quantum dots}
 To preserve the coherence of the electron spins, timescales where the electrons are separated have to be shorter than the decoherence time, $T_2^* = \SI{12.2}{ns}$ (see Fig. 2b). A precise control of the delay between the two-electron transfer is therefore required. It is implemented by triggering the transfer process with nanosecond pulses which load the electrons in the moving potential\cite{hermelin_electrons_2011, bertrand_injection_2016, takada_sound-driven_2019}. Because of the Coulomb interaction, a smaller pulse amplitude is required to send the first electron and a two-pulse procedure permits to control the delay between the electron transfers. The shortest delay is obtained when the two pulses completely overlap and is estimated to be limited by the pulse rise time, equal to $\SI{0.5}{ns}$. Thus, the two electrons are never transferred in the same moving quantum dot. To characterize the efficiency of the time-resolved two-electron transfer, we realize a pump-probe experiment by adding a pulse excitation on the reception dot (see the pulse sequence in Fig. 3a). In this way, electron catching is prevented until a voltage step on gate LP1 is applied. By varying the time delay of this step with respect to the first sending pulse, we can resolve the arrival time of each electron. If this step occurs after the electron arrival time, the catching process does not occur. Figure 3b shows the average charge caught as a function of the sending and catching pulse delays. Three different charge regions are observed and correspond to the catching voltage step happening before, between or after the arrival of the two electrons, leading to the capture of 2, 1 or 0 electrons in the receiver dot, respectively. This two-electron transfer procedure is efficient, with a probability to send both electrons above $95 \pm \SI{1}{\%}$. In addition, the probability for the two electrons to be injected with the intended delay reaches $86 \pm \SI{2}{\%}$ of the successful sendings. From the data presented in Fig. 3, we therefore conclude that we are able to control the electron separation time by adjusting the time delay $\Delta t$ between the two injection voltage pulses with high efficiency. 

\begin{figure*}[!htp]
\includegraphics[width=162mm]{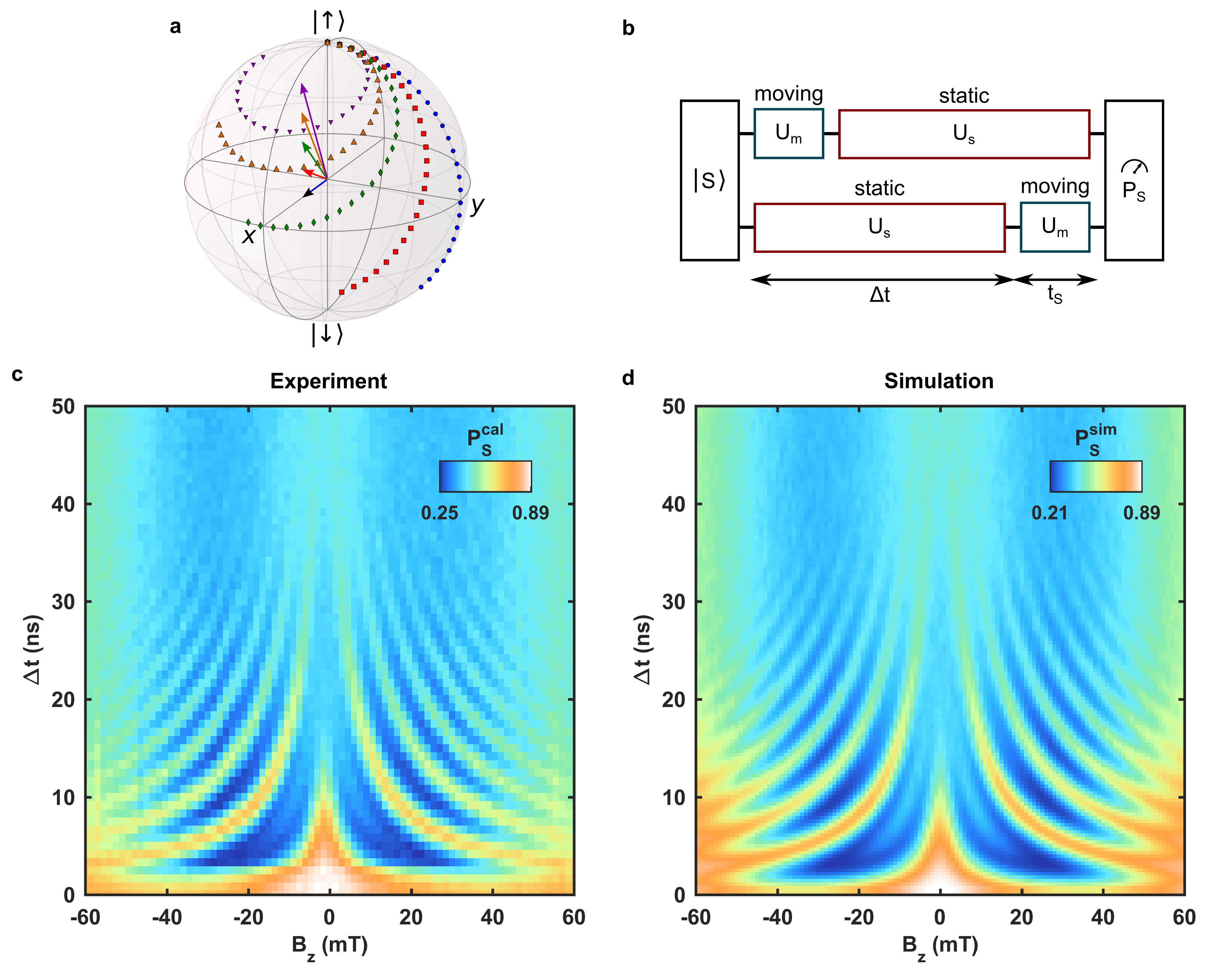}
\caption{ \textbf{Two-electron-spin quantum interferences}
\textbf{a,} Illustration of the spin-orbit driven single-spin rotations for an external magnetic field (orthogonal to the 2DEG plane) of $B_z = 0, 12.5, 25, 37.5 \text{ and }\SI{50}{mT}$ (blue to indigo) and assuming $B\txtind{SO} = \SI{25}{mT}$ and $t\txtind{S} = \SI{2.1}{ns}$. 
\textbf{b,} Scheme of the sequence seen by the two electron spins during transport. The spin-orbit interaction drives identical single-spin rotations on each electron during its displacement. Between these two single-spin operations, the parallel spin states accumulate opposite phases $2\phi = \pm \omega_z \Delta t$ with respect to the anti-parallel states. 
\textbf{c,} Measured singlet spin probability as a function of the sending delay and the external magnetic field. The observed quantum interferences are due to the coherent Larmor precession of the parallel spin states $\uu$ and $\dd$ when the two electrons are static and separated by $\SI{6}{\micro m}$. The oscillations are fitted with a frequency $\omega_z = g \mu B_z / \hbar$, with $g=-0.425 \pm 0.01$. Decoherence towards large separation time is attributed to the hyperfine interaction with the fluctuating nuclear spin environment, acting as an additional random magnetic field of standard deviation $\sim \SI{2}{mT}$ during the separated phase\cite{merkulov_electron_2002}. Each pixel corresponds to 10000 single-shot realizations. 
\textbf{d,} Numerical simulation, performed with the parameters $B\txtind{SO} = \SI{25}{mT}$ and $g=-0.425$ and using 100 repetitions for each pixel. The decoherence mechanisms are reproduced by adding for each repetition a Gaussian nuclear magnetic field of standard deviation of $\SI{2}{mT}$ on each side of the structure and a perturbed electron trajectory. To take into account the charge injection uncertainty, an exponential noise of width \SI{0.5}{ns} is added to the sending delay. The full model is discussed in the Suppl. Inf. 4.
}
\label{fig:interferences_2D}
\end{figure*}

\subsection*{Coherent rotation induced by displacement}
To achieve single-spin coherent manipulation we exploit the strong spin-orbit coupling in GaAs combined with the high velocity of the displacement induced by the SAW. The resulting effective magnetic field $B\txtind{SO}$ induced by spin-orbit interaction therefore overcomes the Overhauser field $B\txtind{HF}$ experienced by the electron during the transfer ($B\txtind{SO} / B\txtind{HF} \approx 15$). Along the transfer sequence, $B\txtind{SO}$ is turned on non-adiabatically when the electron is injected into a moving quantum dot. It is aligned in the plane of the 2DEG and contributing to the total magnetic field experienced by the electron with the orthogonal external magnetic field $B_z$. As a consequence, each displaced electron is coherently rotated by a characteristic angle $\theta$ about an axis $\vec{u}$, both dependent on the strength of the spin-orbit interaction along the electron path, the external magnetic field and the channel length. We designed the channel such that $\theta$ is around $3\pi/4$ at $B_z = \SI{0}{mT}$ and permits to cover a large variety of possible two-electron spin states when varying the magnetic field $B_z$. The resulting unitary transformation $\textrm{U}\txtind{m}$ is expressed in the spin basis of one electron ($\ket{\uparrow}$;$\ket{\downarrow}$) as:    
\begin{equation*}
    \begin{aligned}
    \textrm{U}\txtind{m} &= 
    \exp \left( \frac{-i g \mu\txtind{B}}{2 \hbar} \left( B\txtind{SO} \, \sigma_x + B_z \, \sigma_z \right) t\txtind{S} \right)
    \\ &=
    \begin{pmatrix}
        \Uuu & -i\Uud\\
        &\\
        -i\overline{\Uud} & \overline{\Uuu}
    \end{pmatrix},
    \end{aligned}
\end{equation*}
\begin{equation*}
    \begin{aligned}
    \text{with }
    & \Uuu = \cos \left( \frac{\omega\txtind{tot} t\txtind{S}}{2} \right) + i \frac{\omega_z}{\omega\txtind{tot}} \sin \left( \frac{\omega\txtind{tot} t\txtind{S}}{2} \right), \\
    & \Uud = \frac{\omega\txtind{SO}}{\omega\txtind{tot}} \sin \left( \frac{\omega\txtind{tot} t\txtind{S}}{2} \right),\\
    \text{and } &\omega\txtind{tot} = \frac{g \mu_B}{\hbar} \sqrt{B\txtind{SO}^2 + B_z^2}.
    \end{aligned}
\end{equation*}

Examples of single-spin trajectories for different external magnetic fields are shown in Fig. 4a. As $B_z$ is increased, the single-spin rotation leads to a smaller exploration of the Bloch sphere, the rotation axis converging towards the pole. At $B_z=B\txtind{SO}=\SI{25}{mT}$, the unitary transformation $\textrm{U}\txtind{m}$ is close to an Hadamard gate, mapping each pole to two opposite points of the equator. 

\subsection*{Two-electron quantum interferences}
When combining the controlled ns-delay transfer protocol with the initialization and spin readout sequences, quantum interferences are observed, with high-contrast oscillations of the singlet population (see Fig. 4c). In addition to the spin rotation induced by the electron motion, the electron spins are experiencing the Larmor precession during the time they are separated. It results in a phase shift between the two spin states dependent on $B_z$, and the sending delay $\Delta t$.
\begin{equation*}
    \begin{aligned}
    \textrm{U}\txtind{s}
    &= \begin{pmatrix}
        e^{+i\phi/2}  & 0\\
        0 & e^{-i\phi/2}
    \end{pmatrix}\\
    \text{with }
    & \phi = \omega_z \Delta t = \frac{g \mu_B}{\hbar} B_z \Delta t.
    \end{aligned}
\end{equation*}
In the two-electron spin basis, the complete transformation associated with the ns-delay transfer procedure, illustrated in Fig. 4b, is thus:
\begin{equation*}
    \textrm{U}\txtind{tot} = \textrm{U}\txtind{s} \textrm{U}\txtind{m} \otimes \textrm{U}\txtind{m} \textrm{U}\txtind{s}.
\end{equation*}
After recombination, the singlet spin probability is equal to:
\begin{equation*}
    \PS = \left| 1 + |\Uud|^2 \Big( \cos(\omega_z \Delta t) - 1 \Big) \right| ^2.
\end{equation*}
    
For the condition $B_z=B\txtind{SO}=\SI{25}{mT}$ previously discussed, the interferences should exhibit a unity contrast ($|\Uud|^2 = \frac{1}{2}$). In this case, a maximum coherent transfer of population between the parallel and anti-parallel spin basis is achieved. Following this model, we perform the simulation of the experiment and compute the singlet probability as a function of $\Delta t$ and $B_z$, as shown in Fig. 4d. This simulation is in qualitative and quantitative agreement with the data when taking into account the decoherence mechanisms as described in the Suppl. Info. 4. In particular, a spin-orbit length of $l\txtind{SO} = \SI{8.5}{\micro m}$, in good agreement with the literature\cite{hanson_spins_2007,stotz_coherent_2005,sanada_acoustically_2011}, is extracted.
    
\subsection*{Fast coherent spin displacement}
At zero magnetic field, the singlet state is preserved after transfer because of time-reversal symmetry, and thus no oscillations are observed. Up to a \SI{5}{ns} separation time, we measure a high singlet spin probability, which confirms an efficient coherent transfer between the two distant dots. Quantitatively, we obtain a $0.895 \pm 0.003$ fidelity of the singlet transfer at zero time delay and zero magnetic field ($0.824 \pm 0.003$ without the spin to charge fidelity calibration). The small loss of fidelity is not captured by the simulation and is probably due to spin mixing occurring during the injection process between the static and moving quantum dots\cite{bertrand_injection_2016}.
    
As $B_z$ is increased, coherent oscillations of the singlet probability both in magnetic field and sending delay are observed. From the frequency of these oscillations we extract the electron spin g-factor $g = -0.425 \pm 0.01$. As expected, a maximum population transfer from singlet to triplet parallel spin states is observed for $B_z=\SI{22.5}{mT}$ and $\Delta t=\SI{5}{ns}$ with a minimum singlet probability equal to $0.249 \pm 0.005$ ($0.246 \pm 0.005$ without correction).

The coherent oscillations are characterized by a progressive loss of contrast both in magnetic field and time delay. The comparison between the experimental results and the simulation permits to identify the main sources of decoherence. The contrast reduction in time delay is due to the hyperfine interaction when the electrons are separated. The characteristic decoherence time measured at zero magnetic field is similar to those measured for electrons separated in adjacent dots (see Fig. 2b). At a given time delay, the decoherence observed with the external magnetic field points at a mixing process during the shuttling itself. It originates from the potential fluctuations along the electron path, with an effect on the spin coherence proportional to $B_z$\cite{golovach_phonon-induced_2004,huang_spin_2013}. The strength of these potential fluctuations used in the simulation ($\SI{110}{kV/m}$) is consistent with the randomness of the electrical potential induced by the Si dopants used in our heterostructure\cite{nixon_potential_1990}.

\subsection*{Demonstration of the distant entanglement}
For delays $\Delta t$ above $\SI{2.1}{ns}$, between the arrival of the first electron and the departure of the other one, the system consists of an entangled pair of two individual electron spins stored in two quantum dots separated by $\SI{6}{\micro m}$. To demonstrate the preservation of the initial state entanglement, we have interpreted the two-electron interference in the framework of a Bell type of experiment. It indeed consists in preparing the two electron spins in an entangled state, applying single-spin rotations separated by a phase accumulation, and finally measuring the singlet probability after recombination. The protocol provides a large exploration of the Bloch sphere, as both the azimuthal angle and the latitude are varied over wide ranges. This allows to explore a large diversity of two-electron spin-states only permitted by quantum correlations and leads to important population oscillations (see Fig. 4c).

Proof of the entanglement preservation all along the transfer is imprinted in the interference contrast. In the case of an entangled state $\rho$ made of antiparallel spin states, and $|\Uud|^2 = \frac{1}{2}$, the contrast is expected to be equal to $(\rho_{\ud}+\rho_{\du})/2-\mathrm{Re} \left( \rho_{\ud,\du} \right)$. Thus, for a perfect state preparation in singlet the contrast is expected to be $1$, whereas completely mixed antiparallel spin states are characterized by a maximum of $0.5$ oscillation contrast. Therefore, a contrast above $0.5$ is a signature of entanglement as it necessarily implies $(\rho_{\ud}+\rho_{\du})/2+|\rho_{\ud,\du}|>0.5$\cite{bennett_purification_1996, sackett_experimental_2000}. In the experimental data presented in Fig. 4c, the minimum and maximum singlet probabilities for $\Delta t > \SI{2.1}{ns}$ are respectively $0.890\pm0.003$ and $0.249\pm0.005$. It corresponds then to an oscillation contrast of $0.641\pm0.008$, larger than $0.5$. This value of the contrast demonstrates the preservation of entanglement along the transfer process. Concomitantly it proves the creation of remote entanglement at $\SI{6}{\micro m}$ in a semiconductor quantum circuit. Note that even when the fidelity of the spin readout is not taken into account to calibrate the experimental data, we still observe a contrast well above this threshold ($0.575\pm0.008$).

\subsection*{Conclusion}
In this article, we demonstrate that two entangled electron spins can be separated and displaced controllably at the nanosecond timescale. During the transfer, each electron experiences a single-spin coherent rotation under the influence of the spin-orbit interaction, due to the fast electron transfer procedure. The system exhibits spin quantum interferences, which demonstrates the coherent nature of the initial singlet state in the individual electron spin basis. The spin transfer process is highly coherent with a maximum fidelity close to $\SI{90}{\percent}$, and produce highly entangled electron spins separated by $\SI{6}{\micro m}$. In comparison with precedent demonstrations of coherent shuttling\cite{fujita_coherent_2017, flentje_coherent_2017}, the demonstrated displacement allows a qubit coherent motion over $\SI{6}{\micro m}$ at a timescale of $\SI{2.1}{ns}$. When combined with fast coherent exchange of spin between two adjacent electrons, it permits to envision long-range coupling between distant qubits at frequencies above $\SI{100}{MHz}$. Transposing this technique to a non-piezoelectric material such as silicon would need some adaptations. One way could be to deposit a piezoelectric material on certain parts of the substrate, but fabrication-wise an easier method would be to generate the moving potential by a set of gates along the channel, oscillating with the appropriate phase difference.

\subsection*{Acknowledgements}
We would like to thank B. Bertrand, M. Nurizzo, M. Vinet and X. Hu for enlightening discussions. We acknowledge support from the technical poles of the Institut N\'eel, and in particular the Nanofab team who helped with the sample realization, as well as P. Perrier, G. Pont, H. Rodenas, E. Eyraud, D. Lepoittevin, C. Hoarau and C. Guttin. A.L and A.D.W acknowledge gratefully the support of DFG-TRR160, BMBF-Q.Link.X 16KIS0867, and the DFH/UFA CDFA-05-06. T.M acknowledges financial support from ERC QSPINMOTION and Quantera Si QuBus.

\subsection*{Author contributions}
B.J fabricated the sample and performed the experiments with the help of P.-A.M, T.M and C.B. B.J and T.M interpreted the data, and wrote the manuscript with input from all the other authors. A.L and A.D.W performed the design and molecular-beam-epitaxy growth of the high mobility heterostructure. All authors discussed the results extensively, as well as the manuscript.

\subsection*{Methods}
\subsubsection*{Sample and setup}
Our device was fabricated using a Si doped AlGaAs/GaAs heterostructure grown by molecular beam epitaxy, with a two-dimensional electron gas (2DEG) $\SI{110}{nm}$ below the crystal surface which has a carrier mobility of $\SI{9.1e5}{cm^2 V^{-1} s^{-1}}$ and an electron density of $\SI{2.79e11}{cm^{-2}}$. It is anchored to the cold finger, which is in turn mechanically attached to the mixing chamber of a homemade dilution refrigerator with a base temperature of $\SI{60}{mK}$. It is placed at the center of a superconducting solenoid generating the static out-of-plane magnetic field. Quantum dots are defined and controlled by the application of negative voltages on Ti/Au Schottky gates deposited on the surface of the crystal. Homemade electronics ensure fast changes of both chemical potentials and tunnel couplings with voltage pulse rise times approaching $\SI{100}{ns}$ and refreshed every $\SI{16}{\micro s}$.

A Tektronix 5014C arbitrary waveform generator with a typical channel voltage rise time ($\SI{20}{\%}$ - $\SI{80}{\%}$) of $\SI{0.9}{ns}$ is used to rapidly change the LP1 and RP1 gate voltages. The charge configurations are read-out on each side by two local electrometers (sensing dots) connected to a reflectometry setup of resonant frequencies $\SI{197}{MHz}$ (left) and $\SI{136}{MHz}$ (right). The electrometer is tuned so that the depth of each resonance depends on the conductance through the sensing dot, which in turn is sensitive to the charge occupancy of the neighbor double quantum dot. After demodulation and filtering, each electrometer signal is acquired by a National Instruments analog-to-digital converter with a $\SI{100}{kHz}$ bandwidth.

To shuttle electrons across the channel, a $\SI{100}{ns}$ microwave burst at the resonant frequency of the inter-digital transducer ($\SI{2.79}{GHz}$) is applied by a Rohde \& Schwarz SMA100A signal generator, with a power of $\SI{18}{dBm}$ on the sample. Because of the IDT geometry, the SAW burst has a $\SI{25}{ns}$ ramp-up phase followed by a $\SI{75}{ns}$ maximum amplitude plateau and a $\SI{25}{ns}$ ramp-down phase. As explained in the main text, we controllably inject two electrons when the SAW maximum amplitude reaches the right double-dot, $\SI{740}{ns}$ after the pulse generation.

\subsubsection*{Data analysis and simulation}
In order to remove from the spin analysis the few occurrences ($\SI{4}{\%}$) when the two electrons are not caught in the receiver dot, we measure the charge configuration on the left dot twice: before and after spin-to-charge conversion. From the first signal we select only the successful charge transfers, and from the second we infer the singlet probability. This spin-to-charge conversion and its fidelity, together with the spin initialization, are covered in the Suppl. Inf. 3.

The numerical simulation presented in Fig. 4d is described in details in the Suppl. Inf. 4.

\bibliographystyle{naturemag}
\bibliography{biblioPaper_2}

\iftrue
\newpage
\renewcommand{\figurename}{Supplementary Figure}
\setcounter{figure}{0}

\begin{figure*}[!htp]
\includegraphics[width=120mm]{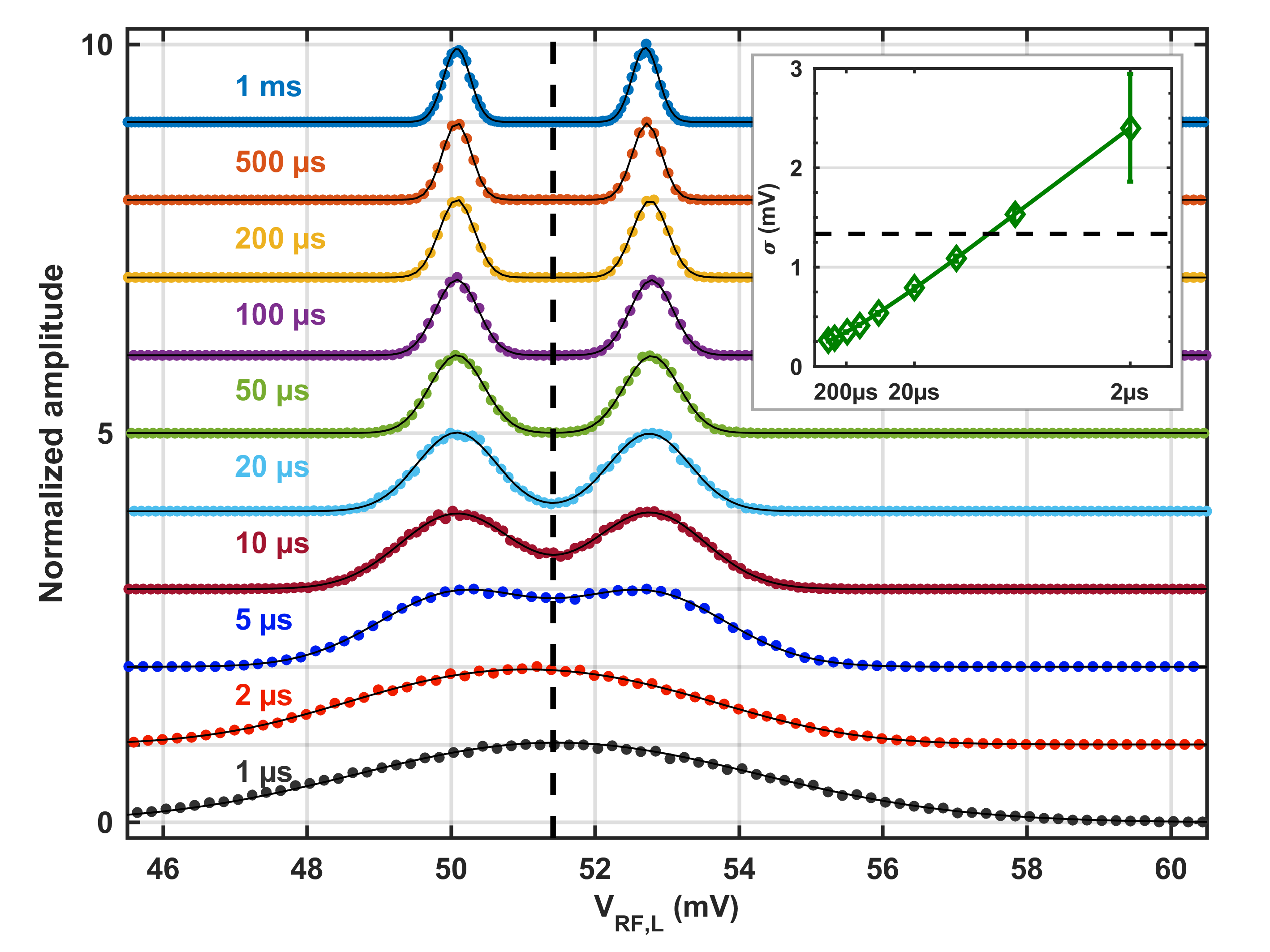}
\caption{ \textbf{Charge sensing and measurement bandwidth.}
Histogram of the signal obtained by loading successively 1 or 2 electrons in the left quantum dot as a function of the integration time. The curves have been normalized and offset for clarity, and fitted with a double Gaussian distribution (solid black line). The dashed black line represents the threshold used for charge assignment, positioned in the middle of the two peaks. \textbf{Inset,} Mean standard deviation of the two electron Gaussian peaks as a function of $1/\sqrt{t\txtind{meas}}$. The noise is inversely proportional to the square root of the integration time, and overcomes the threshold for $t\txtind{meas}<\SI{6}{\micro s}$ (SNR$=1$). For $t\txtind{meas}>\SI{30}{\micro s}$, the charge readout fidelity is greater than $\SI{95}{\%}$.
}
\label{figS1}
\end{figure*}

\subsection*{Supplementary 1 : Charge sensing and measurement bandwidth}
\label{sec:reflecto_bandwidth}
To probe the charge state of our double quantum dots, we use two electrometers (sensing dots), each connected to a resonant circuit ($\SI{136}{}$ and $\SI{197}{MHz}$). The reflected signal is amplified at $\SI{4}{K}$ and at room temperature, demodulated with each carrier frequency, and low-pass filtered at \SI{1}{MHz}. For this bandwidth characterization, we load either one or two electrons in the left double quantum dot and record the signal obtained at the same measurement point using a LeCroy 8Zi oscilloscope. Histograms of the signal acquired for different values of integration time are plotted in Fig. S1.

We observe two Gaussian peaks separated by \SI{2.67}{mV}, corresponding to the two charge states. The noise is proportional to the square root of the measurement speed, with a signal to noise ratio SNR$ = 1$ for \SI{6}{\micro s}, corresponding to the bandwidth of the room temperature amplifier. The error on the charge assignment drops below \SI{5}{\%} for an acquisition time of \SI{30}{\micro s}. 

For the results presented here, since the rest of the measurement sequence takes about \SI{500}{\micro s} per single-shot realization, we increased the readout time to \SI{200}{\micro s} and used a National Instrument acquisition card with a sampling rate of \SI{100}{kHz}.


\begin{figure*}[!htp]
\includegraphics[width=160mm]{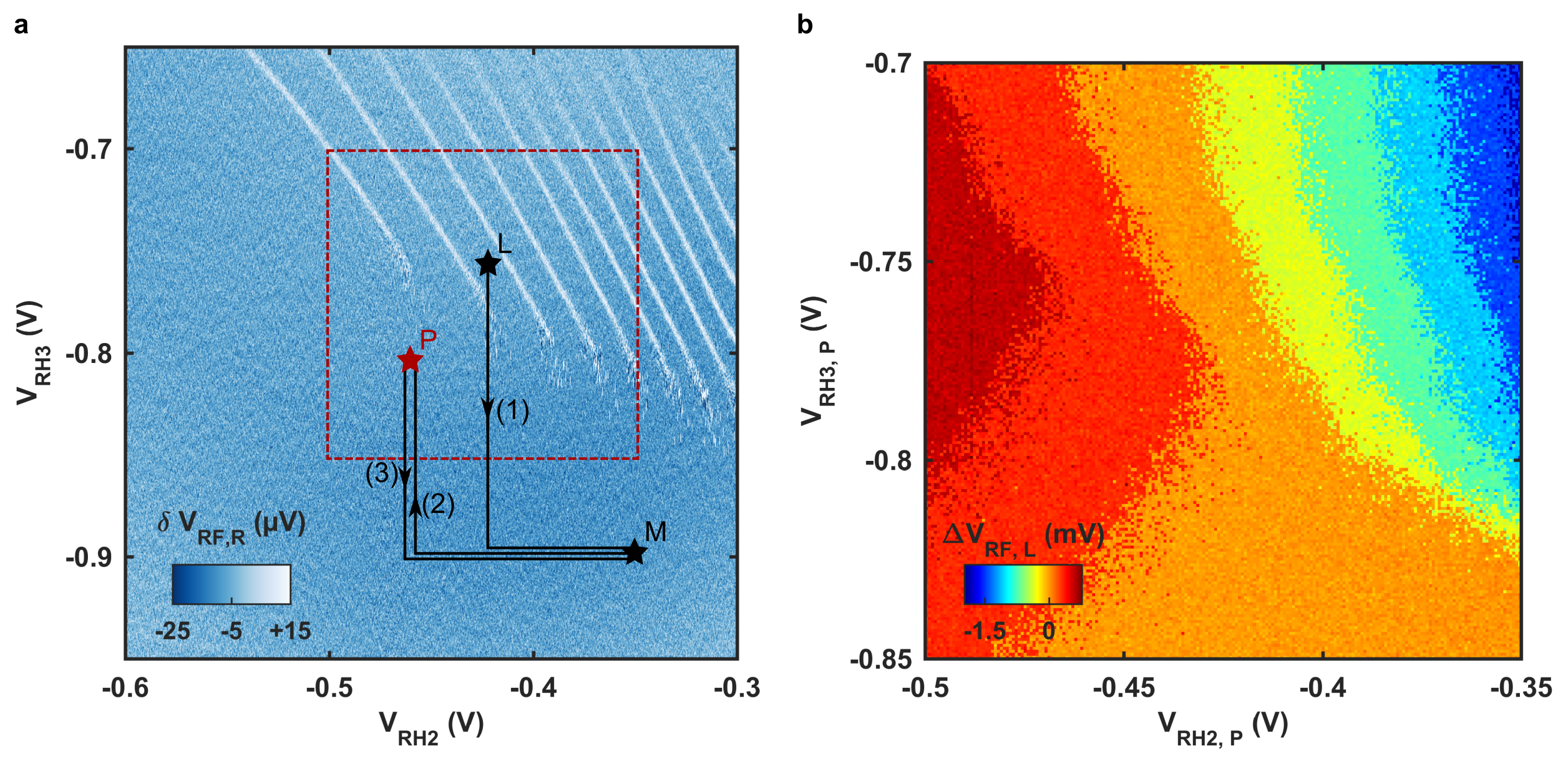}
\caption{ \textbf{Isolated regime.}
\textbf{a,} Scheme of the pulse sequence. Two electrons are loaded at point L, and the signal is recorded for \SI{5}{ms} at point M. We then pulse at a point P for \SI{1}{ms}, and come back to M for another \SI{5}{ms} measurement. 
\textbf{b,} Signal difference before and after the pulse, as a function of the position of the point P, each pixel corresponding to a single-shot realization. In the upper part of the figure, the system is connected to the reservoir and the obtained charge states match perfectly the transitions seen in the charge stability diagram. In the lower part of the diagram (orange area), we can keep our two electrons isolated from the reservoir for the duration of the experiment. This ability to turn on and off the coupling to the reservoir is mandatory to avoid accidental charge exchange via the SAW potential modulation.
}
\label{figS2}
\end{figure*}

\subsection*{Supplementary 2 : Isolated regime}
\label{sec:isolated_regime}
To shuttle a controlled number of electrons or read the charge state on each side easily, we isolate our double dots from the reservoirs, and only increase the coupling with these reservoirs when we need to exchange electrons (to load or perform spin to charge conversion). In the stability diagram shown in Fig. S2a, we can see that the charge transition lines fade out as RH3 becomes more negative, trapping the electrons in the structure. To check that we can indeed hold electrons isolated from the reservoir for the duration of the experiment, we load two electrons in the right double quantum dot and record the signal for \SI{5}{ms} at an isolated position M. We then pulse RH1 and RH3 for \SI{1}{ms} and record again the signal at the same position. 

We plot in Fig. S2b the signal difference as a function of the pulse amplitude in RH1 and RH3. This figure is obtained single-shot for every pixel, and allows us to identify the areas where charges were exchanged with the reservoir, and more importantly the large (orange) area where the two electrons remained isolated in the structure. For the results presented in the main text, the sending and catching position were always far in this isolated regime, in order to be protected from the SAW potential modulation.


\begin{figure*}[!htp]
\includegraphics[width=160mm]{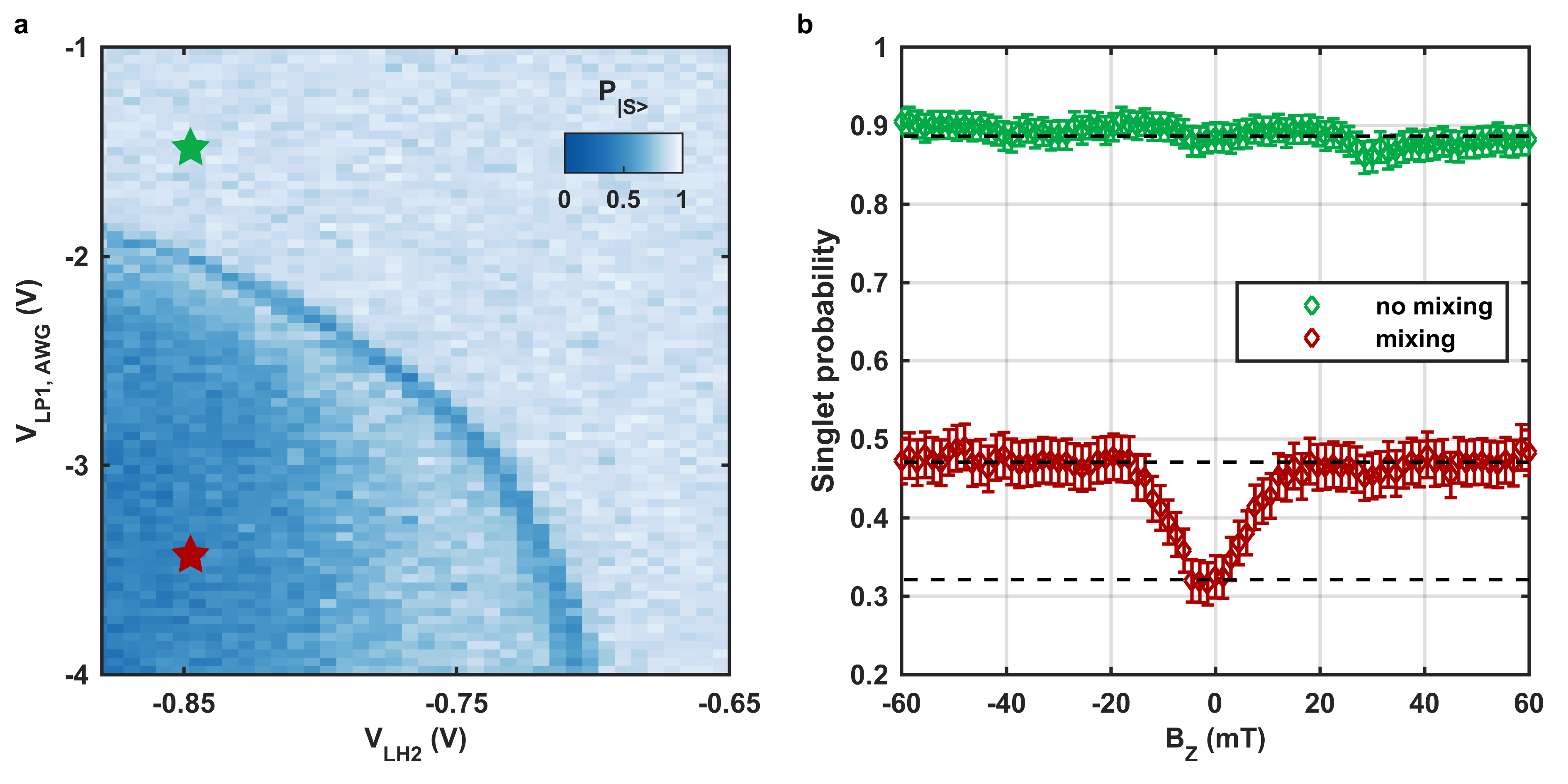}
\caption{ \textbf{Spin initialization and readout fidelity.}
\textbf{a,} Two-electron spin map on the left double dot. The vertical (resp. horizontal) axis is equivalent to the detuning $\epsilon$ (resp. the tunnel coupling $t_c$) between the two dots. On the bottom-left part, we see $S$-$T_0$ mixing in the (1,1) configuration when the exchange coupling is small. The curved line corresponds to $S$-$T_+$ mixing. The two marked positions are used for the spin readout calibration.
\textbf{b,} Measured singlet spin probability as a function of the magnetic field, using a \SI{200}{ns} AWG pulse to the two marked positions. In the (2,0) configuration, the singlet spin is always preserved. In the (1,1) configuration, mixing with the $\ket{T_0}$ state occurs at high magnetic field and the expected singlet population is $\SI{50}{\%}$. For $|B_z|<\SI{20}{mT}$, the triplet degeneracy is not lifted and all spin states are mixed, reducing the singlet population to $\SI{33}{\%}$\cite{merkulov_electron_2002}. From those three asymptotes (dashed black lines) we can extract our spin loading and readout fidelities. 
}
\label{figS3}
\end{figure*}

\subsection*{Supplementary 3 : Spin initialization and readout fidelity}
\label{sec:spin2charge_calibration}
To initialize a singlet spin state, we could load two electrons and wait for their decay to the ground state, which would typically take $\sim \SI{20}{ms}$ and greatly impact our measurement speed. We instead load electrons directly in the ground state by exploiting the \SI{1}{\micro s} rising time of our voltage pulses. We lower the quantum dot potential in a configuration where the coupling to the reservoir is large, which allows the ground state (singlet state) to be populated as soon as it crosses the Fermi energy, while the triplet states are still higher in energy and thus unreachable. This procedure guarantees a high-fidelity singlet initialization in \SI{32}{\micro s}.

As mentioned in the main text, we use a tunnel-rate selective spin readout mechanism. The coupling to the reservoir is momentarily increased, using a \SI{16}{\micro s} voltage pulse. All spin states are above the Fermi energy, but because the triplet states are more coupled to the environment, their tunnel-out rate is greater. We carefully tune the voltage pulse amplitude to reach the maximum spin to charge contrast, using spin mixing in the (1,1) area (see main text) to prepare either a pure singlet state or an equiprobable $S$-$T_0$ mixture. The maximum visibility is \SI{95}{\%} for the left side of the structure and \SI{77}{\%} for the right side.

To calibrate the initialization and readout fidelities, we again used the (1,1) spin mixing and recorded the measured singlet probability as a function of the magnetic field with and without mixing pulse. We call $\alpha$ the probability to measure a singlet state when the system is really in a triplet state, $\beta$ the probability to measure a triplet state instead of a singlet state, and $\gamma$ the singlet initialization errors. The expected singlet probabilities are :
\begin{equation*}
\begin{split}
P_{100} & = (1-\gamma)(1-\beta) + \gamma \alpha \quad \text{for }t\txtind{mix} = \SI{0}{ns}\\
P_{50} & = \frac{1-\beta}{2} + \frac{\alpha}{2} \quad \text{for }t\txtind{mix} = \SI{200}{ns} \text{ and }|B_z| \gg \SI{2}{mT}\\
P_{33} & = \frac{1-\beta}{3} + \frac{2\alpha}{3} \quad \text{for }t\txtind{mix} = \SI{200}{ns} \text{ and }B_z \approx \SI{0}{mT}.
\end{split}
\end{equation*}
From the data of Fig. S3 we extract the parameters for the left spin to charge conversion
\begin{equation*}
\begin{split}
\alpha & = 0.022\\
\beta & = 0.081\\
\gamma & = 0.036.
\end{split}
\end{equation*}
In the main text, the calibrated probability is computed using these parameters as
\begin{equation*}
\PS\txtexp{cal} = \frac{\PS\txtexp{meas}\alpha}{1-\alpha-\beta}.
\end{equation*}
We obtain a readout visibility $(1-\frac{\alpha+\beta}{2}) = \SI{94.9}{\%}$, while the initialization fidelity is $(1-\gamma) = \SI{96.4}{\%}$, limited by the sample electronic temperature.

\subsection*{Supplementary 4 : Simulation}
\label{sec:simulation}
The system is submitted to a perpendicular magnetic field $B_z$, leading to a Zeeman Hamiltonian:
\begin{equation*}
H_z = \frac{1}{2} g \mu_B B_z \sigma_z.
\end{equation*}

During the electron motion at the SAW velocity, the spin-orbit interaction acts as an in-plane magnetic field, perpendicular to the direction of motion $y=\left[\bar{1}10\right]$:
\begin{equation*}
    H\txtind{SO} = \frac{1}{2} g \mu_B B\txtind{SO} \sigma_x,
\end{equation*}
with $B\txtind{SO}$ dependant on the SAW speed $v$ and the spin-orbit length along the $\left[\bar{1}10\right]$ direction $l\txtind{SO}$:
\begin{equation*}
    B\txtind{SO} = \frac{\hbar}{|g|\mu\txtind{B}} \frac{\pi v}{l\txtind{SO}}.
\end{equation*}

After the electron shuttling across the channel length, this interaction drives a single-spin rotation dependent on the channel length and the external magnetic field, noted $\mathrm{U_m}$ and described in the main text. During the static phase, the electron spin precesses around the $z$ axis and accumulates a phase $\phi$. The complete evolution for the two-electron system is thus:
\begin{equation*}
\mathrm{U}\txtind{tot} = \mathrm{Z}_\phi \times \mathrm{U_m} \otimes \mathrm{U_m} \times \mathrm{Z}_\phi.
\end{equation*}

In the simulation whose output is presented in the main text, we try to reproduce the main features of the experimental data. First, we introduce decoherence due to the hyperfine interaction during the static phase, different for each electron as occurring on different sides of the structure. We thus generate for each simulation shot two three-dimensional vectors $\vec{B\txtind{HF}^L}$ and $\vec{B\txtind{HF}^R}$ whose amplitudes follow a normal probability law of mean value \SI{0}{mT} and standard deviation $\sigma\txtind{HF}=\SI{2}{mT}$, this latter value being extracted from the $T_2^*$ measurement in adjacent double quantum dots. No hyperfine interaction is considered during the displacement, as the large SAW speed leads to an important motional narrowing effect. However, we introduce the effect of the potential disorder coming from the Si-doped layer of the heterostructure.

Indeed, the random position of ionized dopants leads to a large potential gradient $\vec{\nabla} V(\vec{r})$ along the electron trajectory, of standard deviation $\sigma_\epsilon = \SI{110}{kV/m}$. Following the derivation of Huang \textit{et al}\cite{huang_spin_2013}, we transform this spatial disorder into a time-varying electrical field in the electron reference frame $\vec{r} = \vec{r_0} + \vec{r\prime}$. Because the confinement imposed by the metallic gates defining the channel ($\SI{4}{meV}$) is much stronger than the one provided by the SAW ($\hbar \omega_d \sim \SI{0.3}{meV}$), we only consider the effect of the potential gradient along the direction of propagation. This fluctuating electrical field leads to an additional magnetic field, orthogonal both to the direction of propagation and to the total magnetic field $\vec{B} = \vec{B_z}+\vec{B\txtind{SO}}$:\cite{golovach_phonon-induced_2004}
\begin{equation*}
\begin{aligned}
&\vec{\delta B} = 2\vec{B} \times \vec{\Omega},\\
&\text{with } \Omega = \frac{-1}{m^* \omega_d^2 l\txtind{SO}} \frac{\partial V}{\partial y} \left( r_0 \right).
\end{aligned}
\end{equation*}

In the simulation, we separate the channel into sections of $\SI{100}{nm}$, which we find to be the typical correlation length of the potential inhomogeneities. We then generate for each section a random electric field, following a normal distribution of standard deviation $\SI{110}{kV/m}$. We compute the effect of the noisy drive on the single-spin rotation $\mathrm{U}$ over the total channel length. We find that the best fitting parameter for the observed decoherence in the system is a SAW-induced confinement of $w_d = \SI{300}{\micro eV}$, which is in good agreement with an independent SAW amplitude estimation for the same input power. In addition, we observe that the large decoherence observed at high magnetic field and small delay can only be explained if we consider a different path for each electron, perhaps due to the Coulomb repulsion during the motion. Thus, we generate a different set of $\vec{\delta B}$ for each electron transfer.

Finally, because of the finite bandwidth of the room-temperature electronics, we expect the voltage pulse used to inject the two electrons (see main text) to exhibit a rising time close to $\sigma_t = \SI{0.7}{ns}$. Because of this effect, we consider that the two electrons are at least separated by $\SI{0.7}{ns}$, even when a single voltage pulse is applied. In addition, we consider an error rate of $\SI{5}{\%}$ when electrons are not injected with the intended delay but instead by the same voltage pulse, leading to a $\SI{0.7}{ns}$ separation. This error rate is in agreement with the single electron transfer fidelity when two electrons are present in the sending dot.

The simulation output, visible in Fig. 4d of the main text and realized with 100 repetitions per pixel, is in great agreement with the experimental data. The parameters used are either extracted from independent measurements on this sample, or in good agreement with similar works\cite{stotz_coherent_2005,huang_spin_2013}.

\fi

\end{document}